\documentclass[12pt]{article}
\input psfig.sty
\input axodraw.sty

\def\sub#1{_{\lower.25ex\hbox{$\scriptstyle#1$}}}

\newskip\zatskip \zatskip=0pt plus0pt minus0pt
\def\matth{\mathsurround=0pt}
\def\lsim{\mathrel{\mathpalette\atversim<}}
\def\gsim{\mathrel{\mathpalette\atversim>}}
\def\atversim#1#2{\lower0.7ex\vbox{\baselineskip\zatskip\lineskip\zatskip
  \lineskiplimit 0pt\ialign{$\matth#1\hfil##\hfil$\crcr#2\crcr\sim\crcr}}}

\def\Moller{M{\o}ller}
\def\sp{s^{'}}
\def\tp{t^{'}}
\def\up{u^{'}}
\renewcommand{\thefootnote}{\fnsymbol{footnote}}

\begin{document} \begin{titlepage} 
\rightline{\vbox{\halign{&#\hfil\cr
&SLAC-PUB-9532\cr
&October 2002\cr}}}
\begin{center} 
\openup4.0\jot
 
{\Large\bf Radiative Corrections to Fixed Target {\Moller} Scattering Including
Hard Bremsstrahlung Effects}
\footnote{Work supported by the Department of 
Energy, Contract DE-AC03-76SF00515}
\medskip

\normalsize 
{\bf \large Frank J. Petriello}
\vskip .2cm
Stanford Linear Accelerator Center \\
Stanford University \\
Stanford CA 94309, USA\\
\vskip .2cm

\end{center} 
\openup2.0\jot
\begin{abstract} 

We present a calculation of the complete ${\cal O}\left(\alpha\right)$ electroweak radiative corrections to 
the {\Moller} scattering process $e^- e^- \rightarrow e^- e^-$, including hard bremsstrahlung contributions.  
We study the effects of these corrections on both the total cross 
section and polarization asymmetry measured in low energy fixed target experiments.  
Numerical results are presented for the experimental cuts relevant for E-158, a fixed 
target $e^- e^-$ experiment being performed at SLAC; the effect of hard bremsstrahlung 
is to shift the measured polarization asymmetry by $\approx +4\%$.  We briefly discuss the 
remaining theoretical uncertainty in the prediction for the low energy {\Moller} scattering 
polarization asymmetry.

\end{abstract}

\renewcommand{\thefootnote}{\arabic{footnote}} \end{titlepage}


\section{Introduction} 

The search for physics beyond the Standard Model (SM) can be pursued in two distinct 
ways: by observing the direct production of particles associated with this new physics, or 
by detecting the indirect influence of these new states on precision measurements.  Since the 
discovery of the top quark in the early 1990s, no new elementary particle has been 
found.  However, many indirect constraints on the SM and its possible extensions 
have been obtained by experiment.  For example, high precision measurements on the $Z$-pole at 
LEP, combined with the values of $M_W$ and $m_t$ measured at the Tevatron, have constrained the 
mass of the SM Higgs boson to satisfy the bound $m_H < 196$ GeV at 95\% CL by studying its 
influence on the radiative corrections to $Z \rightarrow f\bar{f}$ decays~\cite{Abbaneo:2001ix}.  
Similarly, the parameter space available to supersymmetric theories is becoming restricted by 
the absence of CP-violating electric dipole moments which are required in  
supersymmetric models to explain the baryon asymmetry of the universe via electroweak 
baryogenesis~\cite{Chang:2002ex,Pilaftsis:2002fe}.

There are currently several quantities whose measured values do not match their SM predictions.  
The value of the muon $g-2$~\cite{Bennett:2002jb}, ${\rm sin}^2 \left(\theta_W 
\right)$ as derived from the low energy neutrino-nucleon scattering experiment NuTeV~\cite{Zeller:2001hh}, and 
the magnitude of atomic parity violation (APV) in the cesium atom~\cite{Bennett:1999pd} have all 
been recently reported to deviate from their SM prediction by $\gsim 2.5 \, \sigma$.  The 
interpretation of these discrepancies, however, is rendered difficult by our current inability to reliably 
calculate the required hadronic or atomic physics that affects these quantities.  The deviations 
discussed above have been argued to either be less significant or to vanish altogether when the 
relevant contributions are more carefully analyzed~\cite{Melnikov:2001uw}.

These results increase the importance of E-158, a fixed target $e^- e^-$ experiment 
currently being performed at SLAC~\cite{Carr:1997fu}.  This experiment will determine 
${\rm sin}^2 \left(\theta_W \right)$ at a momentum transfer $Q^2 \approx 0.02 \, {\rm GeV}^2$, 
comparable to the value relevant for neutrino-nucleon scattering, by measuring the 
polarization asymmetry $A_{LR} = \left( \sigma_L - \sigma_R \right) / 
\left( \sigma_L + \sigma_R \right)$ in the {\Moller} scattering process $e^- e^- \rightarrow e^- 
e^-$; this reaction is less sensitive to the hadronic physics that complicates the 
previous measurement of ${\rm sin}^2 \left(\theta_W \right)$ by NuTeV, and its interpretation should therefore not be plagued by 
uncontrollable theoretical uncertainties.  The error on the E-158 measurement is expected to reach $\delta A_{LR}/A_{LR} 
\approx \pm 8\%$, corresponding to $\delta \, {\rm sin}^2 \left(\theta_W \right) = \pm 0.0008$, making it 
the most accurate determination of ${\rm sin}^2 \left(\theta_W \right)$ at low momentum 
transfer~\cite{E158talk}.

The study of the electroweak (EW) radiative corrections to {\Moller} scattering was initiated by 
Czarnecki and Marciano~\cite{Czarnecki:1995fw}.  They found that the one-loop corrections 
reduce the tree-level prediction for $A_{LR}$ by $\approx 40 \% $ at low $Q^2$ in a $\overline{MS}$ 
renormalization scheme; the large size of this effect can be traced to the fact that while the tree 
level prediction is proportional to the numerically small electron vector coupling, $g_v = -1/4 
+ {\rm sin}^{2} \left( \theta_W \right) \approx -10^{-2}$, the one-loop result contains quark 
contributions to $\gamma -Z$ mixing not similarly suppressed.  Although the light quark components 
cannot be computed in perturbation theory, they can be related to $e^+ e^- \rightarrow {\rm hadrons}$ 
scattering data; the analysis of~\cite{Czarnecki:1995fw} concludes that these effects can be 
determined with adequate precision.  The virtual electroweak corrections to {\Moller} scattering 
for arbitrary energies were calculated in~\cite{Denner:1998um}, while the pure QED component of $e^- e^- \rightarrow e^- e^-$ was 
computed in~\cite{Shumeiko:1999zd} without experimental cuts.  Neither of these results is sufficient for comparison with the 
E-158 measurement.

In this paper we compute the complete ${\cal O}\left(\alpha\right)$ EW corrections to {\Moller} scattering, 
including hard bremsstrahlung effects, and impose the experimental cuts relevant for the 
E-158 measurement at SLAC.  Our recalculation of the virtual EW corrections is valid for arbitrary
center-of-mass energies and serves as a check of the results in~\cite{Denner:1998um}, with 
which we agree.  We find that the majority of the QED corrections to $A_{LR}$ cancel when 
both virtual and hard photon corrections are added, as anticipated in~\cite{Czarnecki:1995fw,
Denner:1998um}; they contribute only a small $\approx +4 \%$ shift to $A_{LR}$.  This is due to the 
experimental setup of E-158, which treats photons inclusively.  

The paper is organized as follows.  In Section 2 we present our notation and briefly discuss the 
tree-level {\Moller} scattering process.  We describe the calculation of the one-loop electroweak 
corrections in Section 3, and explain the treatment of the hard bremsstrahlung effects in 
Section 4.  We present the appropriate experimental cuts and numerical results in Section 5, 
and conclude in Section 6.

\section{Preliminaries and leading order results}

In this section we present the Born-level predictions for the unpolarized cross section and 
polarization asymmetry for the process $e^- (p_1) e^-(p_2) \rightarrow e^-(q_1) e^-(q_2)$.  The notation 
introduced here will be used throughout the paper.

We first introduce the following Mandelstam invariants:
\begin{eqnarray}
s = (p_1+p_2)^2 \,\, , & & \sp = (q_1+q_2)^2 \,\, , \nonumber \\ 
t = (p_1-q_1)^2 \,\, , & & \tp = (p_2-q_2)^2 \,\, , \nonumber \\ 
u = (p_1-q_2)^2 \,\, , & & \up = (p_2-q_1)^2 \,\, .
\label{mandelstam}
\end{eqnarray}
In the $2 \rightarrow 2$ scattering process described by the Born-level prediction and the virtual 
corrections, the primed invariants are equivalent to their unprimed counterparts: $s = \sp$, $t = \tp$, 
and $u = \up$.  These relations will not hold when we consider the corrections arising from photon 
emission.  We next introduce the electron vector and axial couplings to the $Z$ boson, $g_v = -1/4 +s^{2}_{W}$ 
and $g_a = 1/4$, where $s^{2}_{W} = {\rm sin}^2 \left(\theta_W \right)$ is defined in the on-shell 
renormalization scheme: $s^{2}_{W} = 1 - M_{W}^2 / M_{Z}^2$.  It is convenient to define the following energy-dependent 
effective couplings:
\begin{eqnarray}
c_{LL}\left(x \right) &=& e^2 \left\{ \frac{1}{x}+\frac{(g_v-g_a)^2}{s^{2}_{W} c^{2}_{W} \left(x-M_{Z}^2 
\right)} \right\} \,\, , \nonumber \\
c_{RR}\left(x \right) &=& e^2 \left\{ \frac{1}{x}+\frac{(g_v+g_a)^2}{s^{2}_{W} c^{2}_{W} \left(x-M_{Z}^2 
\right)} \right\} \,\, , \nonumber \\
c_{LR}\left(x \right) &=& e^2 \left\{ \frac{1}{x}+\frac{(g_v^2-g_a^2)}{s^{2}_{W} c^{2}_{W} \left(x-M_{Z}^2 
\right)} \right\}\,\, ,
\label{effcoup}
\end{eqnarray}
where $-e$ is the charge of the electron.  Using these expressions, we can write the squared tree-level 
matrix elements as
\begin{eqnarray}
| {\cal M}^{0}_{LL}|^2 &=& 4\, \left( c_{{{\it LL}}} \left( t \right) +c_{{{\it LL}}} \left( u
 \right)  \right) ^{2}{s}^{2} \,\, , \nonumber \\
| {\cal M}^{0}_{RR}|^2 &=& 4\, \left( c_{{{\it RR}}} \left( t \right) +c_{{{\it RR}}} \left( u
 \right)  \right) ^{2}{s}^{2} \,\, , \nonumber \\
| {\cal M}^{0}_{LR}|^2 &=& | {\cal M}^{0}_{RL}|^2 = 4\, \left( c_{{{\it LR}}} \left( t \right)  
\right) ^{2}{u}^{2}+4\,  \left( c_{{{\it LR}}} \left( u \right)  \right) ^{2}{t}^{2} \,\, .
\end{eqnarray}
The subscripts encode the polarizations of the initial electrons, with $L$ denoting left-handed states 
and $R$ representing right-handed states.  All terms of ${\cal O} (m_{e}^2 /x)$, where $m_e$ is the electron 
mass and $x$ denotes one of the invariants introduced in Eq.~\ref{mandelstam}, have been neglected.  
This is an excellent approximation in the relativistic limit, and is appropriate for the E-158 experimental 
setup.  We next define polarized differential cross sections,
\begin{equation}
\frac{d \sigma_{ij}}{d \Omega} = \frac{| {\cal M}_{ij}|^2}{128 \pi^2 s} \,\, ,
\label{cross}
\end{equation}
and let $\sigma_{ij}$ denote the integration of these expressions over the region under consideration.  We 
remind the reader that the cross section is invariant under boosts along the beam axis if the boundaries of 
integration are chosen consistently in each frame.  The unpolarized cross section and polarization 
asymmetry can now be written as 
\begin{eqnarray}
\sigma_{u} &=& \frac{1}{4} \left\{ \sigma_{LL}+\sigma_{LR}+\sigma_{RL}+\sigma_{RR} \right\} \,\, , 
\nonumber \\
A_{LR} &=& \frac{\sigma_{LL}+\sigma_{LR}-\sigma_{RL}-\sigma_{RR}}{\sigma_{LL}+\sigma_{LR}+\sigma_{RL}+
\sigma_{RR}} \,\, .
\end{eqnarray}

Before discussing the electroweak corrections to these quantities, we comment briefly on our on-shell choice 
of $s^{2}_{W}$.  Numerically, $s^{2}_{W} \approx 0.2216$ in the on-shell scheme, while the $\overline{MS}$ value  
is $s^{2}_{W}\left(M_Z\right)_{\overline{MS}} \approx 0.2312$.  Recalling that the tree-level $A_{LR}$ 
is proportional to $g_v = -1/4 
+ {\rm sin}^{2} \left( \theta_W \right)$, we see that the choice of the on-shell renormalization scheme 
tends to increase the tree-level asymmetry; the value of $A_{LR}$ is also affected by our choice of $M_{W}$ 
as an input, rather than $G_{\mu}$ as in~\cite{Czarnecki:1995fw} (this choice is made to facilitate comparison with 
the complete EW virtual corrections computed in~\cite{Denner:1998um}; the additional uncertainty induced by the larger 
experimental error in $M_W$ is small compared to the uncertainty arising from the hadronic contributions to $\gamma -Z $ mixing).  
We will find that the electroweak corrections to the Born-level 
prediction are larger in the on-shell scheme than in the $\overline{MS}$ scheme chosen in~\cite{Czarnecki:1995fw} (this 
enhancement of on-shell radiative corrections over those of the $\overline{MS}$ scheme is familiar from other studies of 
EW physics~\cite{Ferroglia:2001cr}).
This indicates that the prescription of~\cite{Czarnecki:1995fw} is probably a more appropriate parameterization of the tree-level 
prediction; however, since we will primarily study the effects of hard bremsstrahlung on $A_{LR}$, this fact will not 
concern us here.

\section{Virtual corrections}

In this section we study the one-loop electroweak corrections to the {\Moller} scattering process.  We reduce 
all one-loop integrals to the scalar Passarino-Veltman basis~\cite{Passarino:1978jh} using 
FORM~\cite{Vermaseren:2000nd}; the numerical evaluation of the resulting integrals is then performed using 
{\it LoopTools}~\cite{Hahn:1998yk}.  The matrix elements obtained are valid for $\sqrt{s} \gg m_e$.  
Since the resulting expressions are rather lengthy, and have been previously 
discussed in the literature~\cite{Denner:1998um}, we will not present them explicitly in this manuscript; they are 
available upon request from the author.  We use dimensional regularization to regulate ultraviolet divergences, and an 
anticommuting $\gamma_5$; this is appropriate since no ABJ anomalies are present~\cite{Jegerlehner:2000dz}.

\subsection{QED corrections}

\begin{figure}
\begin{picture}(350,120)(10,20)


\ArrowLine(20,130)(50,100)
\ArrowLine(50,100)(80,130)
\Photon(30,120)(70,120){4}{3}
\Photon(50,100)(50,60){4}{3}
\ArrowLine(20,30)(50,60)
\ArrowLine(50,60)(80,30)
\put(45,133){$\gamma$}
\put(60,80){$\gamma , Z$}


\ArrowLine(115,30)(155,30)
\ArrowLine(155,30)(195,30)
\ArrowLine(195,30)(235,30)
\ArrowLine(115,100)(155,100)
\ArrowLine(155,100)(195,100)
\ArrowLine(195,100)(235,100)
\Photon(155,30)(155,100){4}{3}
\Photon(195,30)(195,100){4}{3}
\put(140,65){$\gamma$}
\put(205,65){$\gamma$}


\ArrowLine(265,30)(305,30)
\ArrowLine(305,30)(345,30)
\ArrowLine(345,30)(385,30)
\ArrowLine(265,100)(305,100)
\ArrowLine(305,100)(345,100)
\ArrowLine(345,100)(385,100)
\Photon(305,30)(305,100){4}{3}
\Photon(345,30)(345,100){4}{3}
\put(290,65){$\gamma$}
\put(355,65){$Z$}

\put(390,65){$+$ permutations}

\end{picture}
\caption{Representative diagrams contributing to the one-loop QED corrections to {\Moller} scattering.}
\label{QEDdiags}
\end{figure}
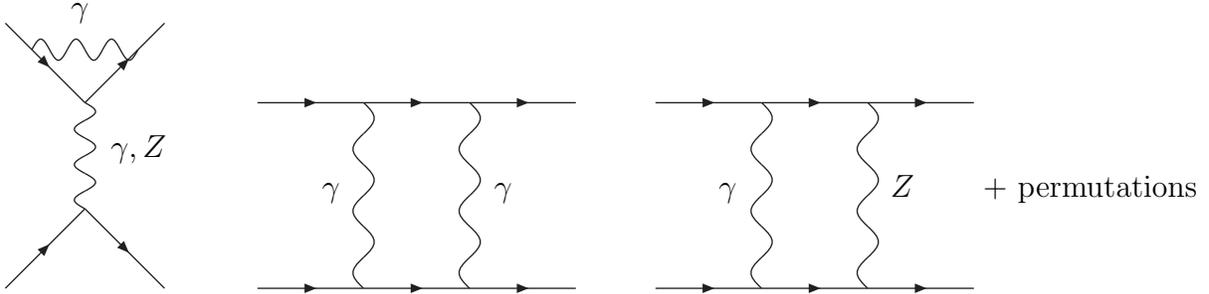

We follow~\cite{Denner:1998um} and define the QED corrections to consist of those diagrams where one photon 
leg has been attached to a tree-level diagram.  A representative sample of the relevant graphs is presented 
in Fig.~\ref{QEDdiags}.  We work in the complete 
on-shell renormalization scheme of~\cite{Denner:kt}, in which self-energy corrections to external fermions do not 
explicitly appear.

These diagrams contain both infrared (IR) and ultraviolet (UV) divergences.  We regulate the IR divergences 
with a finite photon mass $\mu$; these cancel against identical divergences appearing in the process 
$e^- e^- \rightarrow e^- e^- \gamma$, where the photon has an energy $k \leq \Delta E \ll \sqrt{s}$.  In this 
soft photon limit, the integral over the photon phase space factorizes, and its effect is 
to multiply the tree-level squared matrix elements by the following correction factor $\Delta_s$:
\begin{equation}
| {\cal M}^{s}_{ij} |^2 = \Delta_s \, | {\cal M}^{0}_{ij} |^2 \,\, ,
\end{equation}
where $\Delta_s$ has been derived in~\cite{Denner:1998um} and is given by 
\begin{equation}
\Delta_s = \frac{\alpha}{\pi} \left\{ 4 \, {\rm ln}\left(\frac{2 \Delta E}{\mu} \right) \left[ 
{\rm ln}\left(\frac{ut}{m_{e}^2 s} \right) -1 \right] - \left[ {\rm ln}\left(\frac{s}{m_{e}^2} \right)
-1 \right]^2 +1 -\frac{\pi^2}{3} +{\rm ln}^{2} \left( \frac{u}{t} \right) \right\} \,\, .
\end{equation}
The appearance of the cutoff energy $\Delta E$ indicates that this expression is frame-dependent.  The 
E-158 experiment measures the fully inclusive cross section $e^- e^- \rightarrow e^- e^- + (n) \gamma$; the 
dependence on $\Delta E$ will therefore cancel from our results when hard bremsstrahlung effects are included, 
and we need not consider this issue.  The UV divergences can be cancelled by the following wavefunction 
renormalization of the electron field: $ \psi_e \rightarrow \left( 1 + 1/2 \, \delta Z_{\psi} \right) 
\psi_e $.  This renormalization introduces counterterms for both the $\gamma \bar{f}f$ and $Z \bar{f} f$ 
vertices.  In the on-shell scheme, $\delta Z_{\psi}$ can be written in terms of Passarino-Veltman (PV) functions 
as 
\begin{equation}
\delta Z_{\psi} = -\frac{\alpha}{4 \pi} \left\{ B_{0}\left( 0,m_e,0 \right) -1 -4 m_{e}^2 \, 
B^{'}_{0}\left(m_{e}^2 ,m_{e}, \mu \right) \right\} \,\, ,
\end{equation}
where $\alpha$ is the fine structure constant, $B_{0}$ is the two-point PV function in the notation 
of~\cite{Denner:kt}, and $B^{'}_{0}$ is the derivative of this function with respect to its first argument.  
$B^{'}_{0}$ is infrared divergent, and we have therefore retained its dependence on the photon mass $\mu$.

Denoting the one-loop QED matrix elements by ${\cal M}^{Q}_{ij}$, we can 
write the QED corrections to $| {\cal M}_{ij}|^2$ as 
\begin{equation}
\delta^{Q}_{ij} = 2 \, {\rm Re} \left\{ {\cal M}^{Q}_{ij} \left( {\cal M}^{0}_{ij} \right)^{*} \right\} + | {\cal M}^{s}_{ij} 
|^2 \,\, .
\end{equation}

\subsection{Weak corrections}

\begin{figure}
\begin{picture}(500,220)(0,80)


\ArrowLine(0,290)(25,265)
\ArrowLine(25,265)(50,290)
\Photon(25,265)(25,235){3}{3}
\Photon(8,282)(42,282){3}{3}
\ArrowLine(0,210)(25,235)
\ArrowLine(25,235)(50,210)
\put(22,288){$Z$}
\put(32,250){$\gamma ,Z$}


\ArrowLine(90,290)(115,265)
\ArrowLine(115,265)(140,290)
\Photon(115,265)(115,235){3}{3}
\Photon(98,282)(132,282){3}{3}
\ArrowLine(90,210)(115,235)
\ArrowLine(115,235)(140,210)
\put(112,288){$W$}
\put(122,250){$\gamma ,Z$}


\ArrowLine(180,290)(190,280)
\ArrowLine(220,280)(230,290)
\ArrowLine(190,280)(220,280)
\Photon(190,280)(205,265){2}{2}
\Photon(220,280)(205,265){2}{2}
\Photon(205,265)(205,235){3}{3}
\ArrowLine(180,210)(205,235)
\ArrowLine(205,235)(230,210)
\put(185,260){$W$}
\put(212,250){$\gamma ,Z$}


\ArrowLine(270,210)(295,210)
\ArrowLine(295,210)(320,210)
\ArrowLine(320,210)(345,210)
\ArrowLine(270,260)(295,260)
\ArrowLine(295,260)(320,260)
\ArrowLine(320,260)(345,260)
\Photon(295,210)(295,260){3}{3}
\Photon(320,210)(320,260){3}{3}
\put(280,235){$Z$}
\put(328,235){$Z$}


\ArrowLine(380,210)(405,210)
\ArrowLine(405,210)(430,210)
\ArrowLine(430,210)(455,210)
\ArrowLine(380,260)(405,260)
\ArrowLine(405,260)(430,260)
\ArrowLine(430,260)(455,260)
\Photon(405,210)(405,260){3}{3}
\Photon(430,210)(430,260){3}{3}
\put(390,235){$W$}
\put(438,235){$W$}


\ArrowLine(0,180)(25,155)
\ArrowLine(25,155)(50,180)
\Photon(25,155)(25,140){3}{1.5}
\ArrowArc(25,130)(10,0,360)
\Photon(25,120)(25,105){3}{1.5}
\ArrowLine(0,80)(25,105)
\ArrowLine(25,105)(50,80)
\put(30,110){$\gamma ,Z$}
\put(0,130){$f$}


\ArrowLine(90,180)(115,155)
\ArrowLine(115,155)(140,180)
\Photon(115,155)(115,140){3}{1.5}
\PhotonArc(115,130)(10,0,360){1.5}{8}
\Photon(115,120)(115,105){3}{1.5}
\ArrowLine(90,80)(115,105)
\ArrowLine(115,105)(140,80)
\put(120,110){$\gamma ,Z$}
\put(90,130){$W$}


\ArrowLine(180,180)(205,155)
\ArrowLine(205,155)(230,180)
\Photon(205,155)(205,140){3}{1.5}
\DashCArc(205,130)(10,0,360){2}
\Photon(205,120)(205,105){3}{1.5}
\ArrowLine(180,80)(205,105)
\ArrowLine(205,105)(230,80)
\put(210,110){$\gamma ,Z$}
\put(165,130){$h,\phi^{\pm},$}
\put(172,120){$u$}

\put(260,130){$+$ permutations $+$ counterterms}

\end{picture}
\caption{Representative diagrams contributing to the one-loop weak corrections to {\Moller} scattering.}
\label{EWdiags}
\end{figure}
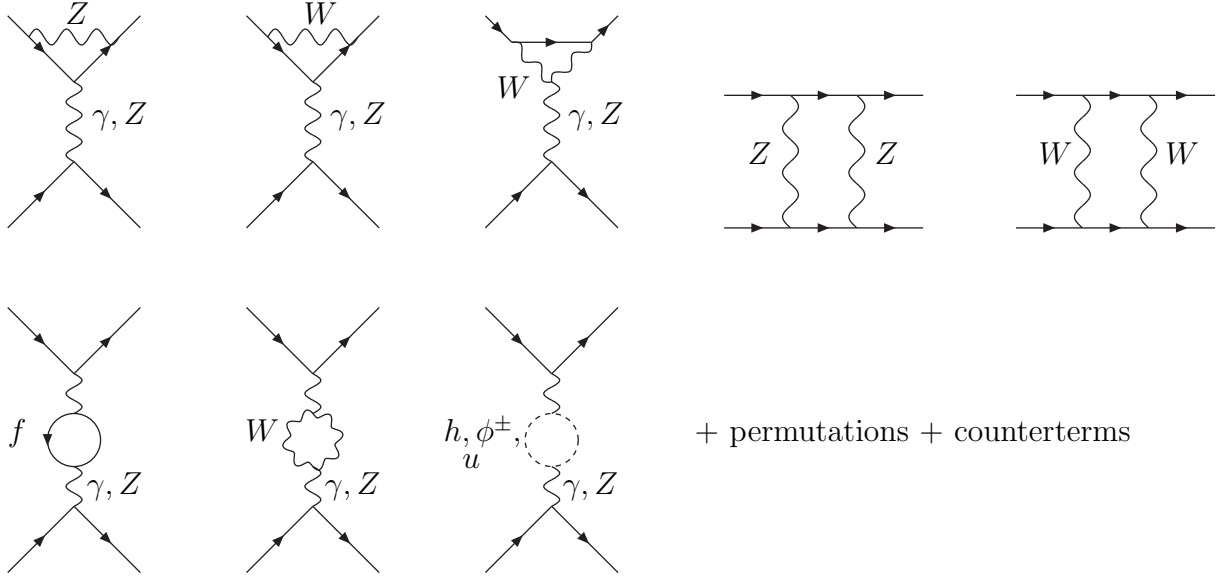

The one-loop weak corrections to {\Moller} scattering consist of the $\gamma$ and $Z$ self-energy graphs,  
$\gamma -Z$ mixing, the remaining vertex and box diagrams, and the appropriate self-energy and vertex counterterms; 
a representative subset of the countributing graphs is presented in Fig.~\ref{EWdiags}.
These contributions contain UV divergences, which must be removed via the full SM renormalization program; we follow 
the on-shell prescription described in~\cite{Denner:kt}.  The self-energy and counterterm graphs 
can be combined and interfered with the Born-level diagrams to yield
\begin{eqnarray}
{\cal M}^{0}_{LL}\left({\cal M}^{s+c}_{LL}\right)^{*} &=& 4\,e^2 \left( \left\{ c_{LL}\left( t \right) 
+ c_{LL}\left( u \right) \right\} \left\{{\cal V}_{LL}\left( t \right) + {\cal V}_{LL}\left( u \right) 
\right\} \right) s^2 \,\, , \nonumber \\ 
{\cal M}^{0}_{RR}\left({\cal M}^{s+c}_{RR}\right)^{*} &=& 4\,e^2 \left( \left\{ c_{RR}\left( t \right) 
+ c_{RR}\left( u \right) \right\} \left\{{\cal V}_{RR}\left( t \right) + {\cal V}_{RR}\left( u \right) 
\right\} \right) s^2 \,\, , \nonumber \\ 
{\cal M}^{0}_{LR}\left({\cal M}^{s+c}_{LR}\right)^{*} &=& {\cal M}^{0}_{RL}\left({\cal M}^{s+c}_{RL}\right)^{*} = 
4\,e^2 \left( c_{LR}\left( t \right) {\cal V}_{LR}\left( t \right) u^2 
+ c_{LR}\left( u \right) {\cal V}_{LR}\left( u \right) t^2 \right) \,\, , 
\end{eqnarray}
where the ${\cal V}_{ij} \left( x \right)$ are given by
\begin{eqnarray}
{\cal V}_{LL} \left( x \right) &=& \left\{ 2 \, \left(\delta Z_e +
\delta Z_L \right)\left[ \frac{1}{x}+\frac{g_{L}^2}{x-M_{Z}^2 } \right] + \frac{g_L}{x-M_{Z}^2} \, \frac{\delta s_{W}}
{s^{2}_{W}c^{3}_{W}}   +\frac{g_{L}^{2} \, \delta M_{Z}^{2}}{\left(x-M_{Z}^{2}\right)^2}  \right. \nonumber \\ 
& & \left. -\frac{\Sigma^{AA}\left( x \right)}{x^2} -2 \, \frac{g_{L}\, \Sigma^{AZ}\left( x \right)}{x \left(x-M_{Z}^2 
\right)} - \frac{g_{L}^{2} \, \Sigma^{ZZ} \left( x \right)}{\left(x-M_{Z}^{2} \right)^{2}} \right\} \,\, ,
\nonumber \\
{\cal V}_{RR} \left( x \right) &=& \left\{ 2 \, \left(\delta Z_e +
\delta Z_R \right)\left[ \frac{1}{x}+\frac{g_{R}^2}{x-M_{Z}^2 } \right] + 2\, \frac{g_R}{x-M_{Z}^2} \, 
\frac{\delta s_{W}} {c^{3}_{W}} +\frac{g_{R}^{2} \, \delta M_{Z}^{2}}{\left(x-M_{Z}^{2}\right)^2} \right. 
\nonumber \\ & & \left. -\frac{\Sigma^{AA}\left( x \right)}{x^2}  -2 \, \frac{g_{R}\, \Sigma^{AZ}\left( x 
\right)}{x \left(x-M_{Z}^2 \right)} - \frac{g_{R}^{2} \, \Sigma^{ZZ} \left( x \right)}{\left(x-M_{Z}^{2} \right)^{2}} 
\right\} \,\, , \nonumber \\
{\cal V}_{LR} \left( x \right) &=&  \left\{ \left(2 \, \delta Z_e 
+\delta Z_L +\delta Z_R \right) \left[ \frac{1}{x} +\frac{g_L g_R}{x-M_{Z}^2} \right]  +\frac{1}{x-M_{Z}^2} 
\left[\frac{g_R}{2 \, s_{W} c_{W}} + \frac{g_{L}\, s_{W}}{c_{W}} \right] \frac{\delta s_{W}}{s_{W}c^{2}_{W}}  
\right. \nonumber \\ & & \left.
+ \frac{g_{L} g_{R} \, \delta M_{Z}^{2}}{\left( x-M_{Z}^{2} \right)^{2}} - \frac{\Sigma^{AA} \left( x \right)}
{x^2} -  \frac{\left(g_{L}+g_{R}\right) \Sigma^{AZ}\left( x \right)}{x \left( x-M_{Z}^{2} \right)} 
-\frac{g_L g_R \, \Sigma^{ZZ} \left( x \right)}{\left( x-M_{Z}^{2} \right)^{2}} \right\} \,\, .
\label{counterterms}
\end{eqnarray}
We have introduced the abbreviations $g_L = \left(g_v - g_a \right)/s_{W}c_{W}$ and $g_R = \left(g_v 
+ g_a \right)/s_{W}c_{W}$ for the left and right-handed electron couplings.  The forms of the self-energy 
insertions $\Sigma^{AA} \left( x \right)$, $\Sigma^{AZ} \left( x \right)$, and $\Sigma^{ZZ} \left( x \right)$, 
as well as expressions for the counterterms $\delta Z_e$, $\delta Z_L$, $\delta Z_R$, $\delta s_{W}$, and 
$\delta M_{Z}^2$ in terms of these functions, can be found in~\cite{Denner:kt}; note that the pure QED component 
of Eq.~\ref{counterterms} has been considered separately in the previous subsection.

The momentum transfers relevant for the E-158 measurement are $|t|,|u| \approx 0.02 \, {\rm GeV}^2$.  At this 
energy scale, the light quark contributions to the $\gamma\gamma$ and $\gamma Z$ vacuum polarization functions in 
Eq.~\ref{counterterms} cannot be computed perturbatively; the appropriate degrees of freedom in this regime 
are the low-lying hadronic states, and the required terms must be obtained by 
comparison to experimental data.  We discuss here the determination of these contributions.  We first 
consider the expression $2\, \delta Z_e - \Sigma^{AA} \left( x \right) /x$.  Evaluating the hadronic contributions 
to this quantity in the free quark approximation, and neglecting the quark masses wherever possible, we find 
\begin{equation}
2\, \delta Z_e - \frac{\Sigma^{AA} \left( x \right)}{x} = \frac{\alpha}{3 \pi} \sum_{q} Q_{q}^{2} N_c \left\{
{\rm ln}\left( \frac{|x|}{m_{q}^{2}}\right) -\frac{5}{3} \right\} + \ldots \,\, ,
\end{equation}
where the sum is over the five light quarks and the ellipsis denotes the terms arising from the remaining states.  
This quantity can be recognized as $\Delta_{{\rm had}} \left(x \right) = \alpha_{{\rm had}}\left(0\right)-\alpha_{{\rm had}}\left(x\right)$, 
which can be related to $e^+ e^- \rightarrow 
{\rm hadrons}$ scattering data via a dispersion relation~\cite{Jegerlehner:1991dq}. We use the parameterization 
of $\Delta_{{\rm had}} \left(x \right)$ given in~\cite{Burkhardt:2001xp}, which is a good approximation throughout the entire range 
of $x$.  A similar calculation reveals that 
\begin{equation}
\frac{2 \, \delta Z_e}{t-M_{Z}^2} = \frac{\Delta_{{\rm had}}\left( M_{Z}^2 \right) + \Sigma^{AA}_{q}
 \left( M_{Z}^2 \right) / M_{Z}^2 }{t-M_{Z}^2} + \ldots \,\, ,
\end{equation}
where only the light quark components have been explicitly shown; $ \Delta_{{\rm had}}\left( M_{Z}^2 \right)$ 
can again be obtained from~\cite{Burkhardt:2001xp}, while $\Sigma^{AA}_{q} \left( M_{Z}^2 \right) $ can be 
calculated perturbatively.

Finally, we must discuss the contribution of the $\gamma -Z$ mixing term, $- \Sigma^{AZ}\left( x \right)
/x \left(x - M_{Z}^{2} \right)$.  For $|x|$ much larger than the light quark masses, this quantity can be 
evaluated perturbatively; as mentioned earlier, this condition does not hold in the E-158 experimental setup.  We begin by 
rewriting this piece using
\begin{equation}
\left[ \frac{\Sigma^{AZ}\left( x \right)}{x} - \frac{1}{\epsilon} \right]  + \frac{1}{\epsilon} = 
\frac{\Pi^{\gamma Z}_{MS}\left(x\right)}{x} + \frac{1}{\epsilon} \,\ , 
\label{AZmix} 
\end{equation}
where $\epsilon = 4-D$ is the usual regulator appearing in dimensional regularization, and $\Pi^{\gamma Z}_{MS}$ is the 
$\gamma Z$ vacuum polarization function defined in the $\overline{MS}$ renormalization scheme.  The remaining $1 / \epsilon$ 
pole cancels when the counterterms of Eq.~\ref{counterterms} are combined with the vertex diagram contributions.  
We next set $x=0$ in $\Pi^{\gamma Z}_{MS}$; an estimate of the error induced by evaluating this quantity 
at $ x=0$ rather than at the value $|x| \approx 0.02 \, {\rm GeV}^2$ relevant for the E-158 
experiment was performed in~\cite{Czarnecki:1995fw} by calculating the pion contributions to 
$\Pi^{\gamma Z}\left( x \right)$ in a scalar QED framework.  This effect was found to be negligible, and we ignore it in 
the remainder of our analysis.  Evaluating $\Pi^{\gamma Z}$ in the free quark approximation, we obtain
\begin{equation}
\Pi^{\gamma Z}_{MS}\left( 0 \right) = \frac{\alpha}{2 \pi} \sum_{q} \left\{ N_{c} Q_{q} \left[T_3 -
2 Q_{q} s^{2}_{W} \right] \left( \frac{1}{3} {\rm ln}\left( \frac{m_{q}^{2}}{M_{Z}^{2}}\right)   \right) \right\}
\,\, .
\end{equation}
A dispersion relation analysis of this quantity was performed in~\cite{Czarnecki:1995fw,Marciano:ss}; we use this result 
and replace 
\begin{equation}
\frac{1}{3} \sum_{q} N_{c} Q_{q} \left[T_3 -
2 Q_{q} s^{2}_{W} \right] {\rm ln}\left( \frac{m_{q}^{2}}{M_{Z}^{2}}\right) \rightarrow -6.88 \pm 0.50\,\, .
\end{equation}
It is claimed in~\cite{Czarnecki:1995fw} that an improved treatment of the $e^+ e^- \rightarrow \, {\rm hadrons}$ 
scattering data would reduce the quoted error; an analysis of this type motivated by the needs of a high energy 
$e^- e^-$ program seems to be underway~\cite{Marctalk}.

Denoting the one-loop weak matrix elements by ${\cal M}^{W}_{ij}$, we can 
write the weak corrections to $| {\cal M}_{ij}|^2$ as 
\begin{equation}
\delta^{W}_{ij} = 2 \, {\rm Re} \left\{ {\cal M}^{W}_{ij} \left( {\cal M}^{0}_{ij} \right)^{*} \right\} \,\, .
\end{equation}
The complete one-loop expressions for the matrix elements become
\begin{equation}
|{\cal M}_{ij}|^2 = |{\cal M}^{0}_{ij}|^2 + \delta^{Q}_{ij} + \delta^{W}_{ij} \,\, .
\label{oneloop}
\end{equation}
Our results are in a form valid for all $\sqrt{s} \gg m_e$; we can compare them with the detailed results 
for the unpolarized cross sections and polarization asymmetries given in Tables 1 and 2 
of~\cite{Denner:1998um}.  When we use the same parameters found there, the older parameterization of 
$\Delta_{{\rm had}}\left( x \right)$ given in~\cite{Burkhardt:1995tt}, and the light quark masses 
given in~\cite{Jegerlehner:1991dq}, we find complete numerical agreement to the given accuracy.

\section{Hard bremsstrahlung corrections}

In this section we describe corrections to {\Moller} scattering arising from the process $e^- (p_1) e^-(p_2) 
\rightarrow e^-(q_1) e^-(q_2) \gamma\left(k\right)$, in which the emitted photon has an energy $k > \Delta E$.  We 
first discuss the calculation of the relevant matrix elements, and afterwards present our parameterization of the 
phase space.

In our calculation of the hard bremsstrahlung corrections to {\Moller} scattering, we again work in the 
relativistic limit $\sqrt{s} \gg m_e$; however, as is well known, $m_e$ cannot be neglected in regions of the 
phase space where the emitted photon becomes collinear with one of the electrons.  Similarly, $m_e$ must be retained 
when one of the final-state electrons travels parallel to the beam axis, and the $t$-channel momentum transfer becomes 
of ${\cal O}\left( m_{e}^{2} \right)$.  We first perform the calculation neglecting $m_e$ 
everywhere, and then discuss the required correction factors.  We introduce the following abbreviations for the electron 
propagator denominators connected to an emitted photon:
\begin{eqnarray}
d_1 = \frac{-1}{2p_1 \cdot k} \,\, , & & d_2 = \frac{-1}{2p_2 \cdot k} \,\, , \nonumber \\
d_3 = \frac{1}{2q_1 \cdot k} \,\, , & & d_4 = \frac{1}{2q_2 \cdot k} \,\, .
\label{bremdenoms}
\end{eqnarray}
The expressions for the squared matrix elements $|{\cal M}^{h,a}_{ij}|^2$ computed with $m_e = 0$ everywhere can be 
found in Eq.~\ref{naive} in the Appendix.  Recalling the definition of $c_{ij}\left(x\right)$ presented in 
Eq.~\ref{effcoup}, we observe that the matrix elements contain peaks of the form $ s/ m_{e}^2$ when one of the 
following eight quantities becomes of ${\cal O}\left(m_{e}^2\right)$: $1/d_i,t,u,\tp,\up$, where $i=1,2,3,4$.  The 
first four peaking regions are associated with the emitted photon becoming collinear with any of the four electrons, 
while the remaining four peaks arise when one of the final-state electrons becomes collinear with an initial-state 
electron.  Note that all terms containing $d_{i}^2,1/t^2,1/u^2,1/t^{'2}$, and $1/u^{'2}$ have cancelled, leaving no 
contribution which behaves like $ s^2/ m_{e}^4$.  However, in the peaking regions, terms of the form 
$m_{e}^2 d_{i}^2$, $m_{e}^2 / t^2$, etc. are of ${\cal O}\left(s/m_{e}^{2}\right)$, and must be included; we now 
discuss the calculation of these pieces.

\begin{figure}
\begin{center}
\begin{picture}(100,120)(0,0)


\ArrowLine(0,0)(50,0)
\ArrowLine(50,0)(100,0)
\Photon(50,0)(50,80){4}{3}
\ArrowLine(0,80)(50,80)
\ArrowLine(50,80)(100,80)
\Photon(75,80)(100,105){4}{3}
\put(0,10){$e^- \left(p_2\right)$}
\put(0,65){$e^- \left(p_1\right)$}
\put(65,10){$e^- \left(q_2\right)$}
\put(65,65){$e^- \left(q_1\right)$}
\put(108,95){$\gamma \left(k\right)$}

\end{picture}
\end{center}
\caption{A typical bremsstrahlung diagram; the self-interference of this particular graph neccessitates 
the inclusion of final-state mass effects.}
\label{bremdiag}
\end{figure}
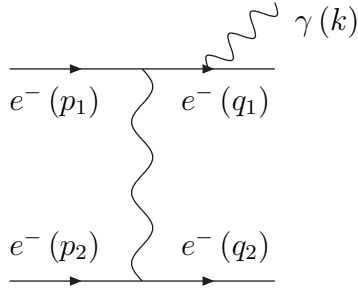
We first consider the mass effects associated with the unpolarized final-state particles.  We focus on 
$e^- \left(q_1\right)$; the discussion proceeds identically for $e^- \left(q_2\right)$.  A typical 
final-state bremsstrahlung diagram is shown in Fig.~\ref{bremdiag}; the self-interference of this diagram, or its 
interference with another diagram where $\gamma\left(k\right)$ is emitted from $e^- \left(q_1\right)$, leads 
to the following term in the squared matrix element:
\begin{eqnarray}
& & \left\{\not\!{q}_1 +\not\!{k} +m_e\right\}\not\!{\epsilon}\left(k\right)u\left(q_1\right)\bar{u}\left(q_1\right)
\not\!{\epsilon}^{*}\left(k\right)\left\{\not\!{q}_1 +\not\!{k} +m_e\right\} \left(d_{3}\right)^{2} \nonumber \\ 
&=& \left\{2 q_1 \cdot \epsilon\left(k\right) + \not\!{k}\not\!{\epsilon}\left(k\right)\right\}u\left(q_1\right)
\bar{u}\left(q_1\right)\left\{2 q_1 \cdot \epsilon^{*}\left(k\right) + \not\!{k}\not\!{\epsilon}^{*}\left(k\right) 
\right\} \left(d_{3}\right)^{2} \,\, ,
\end{eqnarray}
where $\epsilon\left(k\right)$ is the photon polarization vector, $u\left(q_1\right)$ is the Dirac spinor of 
$e^- \left(q_1\right)$, and the equality results from the use of the Dirac equation $\left(\not\!{q}_1 +m_e \right) 
u\left(q_1\right) = 0$.  We now sum over photon polarization vectors and electron spin states; since the electron 
is unpolarized, we replace $u\left(q_1\right)\bar{u}\left(q_1\right) \rightarrow \not\!\!{q}_1 +m_e$.  After standard 
Dirac algebraic manipulations, keeping only $1/d_3$ and $m_{e}^{2}/\left(d_{3}\right)^{2}$ terms, we arrive 
at the expression
\begin{equation}
2 \not\!{q}_{1} d_3 - 4 \, m_{e}^{2} \left( \not\!{k}+\not\!{q}_{1}\right) \left(d_{3}\right)^{2}\,\, .
\end{equation}
We have obtained the required $1/d_3$ and $m_{e}^{2}/\left(d_{3}\right)^{2}$ terms; the remainder of the numerator 
algebra can now be performed in the massless limit. We note that $d_3$ must be evaluated with its full $m_e$ 
dependence in the peaking region; otherwise, $d_3 \rightarrow \infty$, and an unphysical singularity is 
encountered. 

The calculation of initial-state mass effects is similar to the process outlined above.  We find that the 
squared matrix element contains the term
\begin{equation}
\left\{2 p_1 \cdot \epsilon\left(k\right) - \not\!{k}\not\!{\epsilon}\left(k\right)\right\}u\left(p_1\right)
\bar{u}\left(p_1\right)\left\{2 p_1 \cdot \epsilon^{*}\left(k\right) - \not\!{k}\not\!{\epsilon}^{*}\left(k\right) 
\right\} \left(d_{1}\right)^{2} 
\label{initmass}
\end{equation}
when self-interference of diagrams with bremsstrahlung off $e^{-} \left(p_1 \right)$ is considered; a similar 
expression is obtained when studying $e^{-} \left(p_2 \right)$.  Since the initial states can be polarized, we 
must now replace 
\begin{equation}
u\left(p_1\right)\bar{u}\left(p_1\right) \rightarrow \frac{1}{2}\left(\not\!{p}_1 +m_e\right)\left(1+\gamma_5 
\not\!{s}\right) \,\, ,
\end{equation}
where 
\begin{equation}
\not\!{s} = P_{||}\left\{\frac{\not\!{p}_{1}}{m_e}-m_e \frac{\not\!{q}}{p_{1} \cdot q} \right\} \,\, ,
\end{equation}
$P_{||}$ is the degree of longitudinal polarization, and $q$ is an auxiliary vector satisfying $q^2 = 0$, 
$p_1 \cdot q \not\!\!{=} 0$; for mass effects associated with $e^{-} \left(p_1 \right)$, it is convenient to 
choose $q = p_2$.  For the calculation of mass effects arising from $e^{-} \left(p_2 \right)$, we choose $q = p_1$.
After performing standard Dirac algebraic manipulations, again keeping only the leading terms which contribute 
to the squared matrix element, and setting $P_{||} = \pm 1$, we find that Eq.~\ref{initmass} becomes
\begin{equation}
\omega_{\pm}\left\{-2 \not\!{k}d_1 -4\, m_{e}^{2} \left(\not\!{p}_1 - \not\!{k}\right)\left(d_1 \right)^{2} \right\}
\mp \frac{4\,m_{e}^{2}}{s}\left(\omega_+ - \omega_- \right) \left\{ p_1 \cdot p_2 \not\!{k} - p_2 \cdot k 
\left( \not\!{p}_{1} - \not\!{k} \right) \right\} \left(d_1 \right)^{2} \,\, ,
\end{equation}
where $\omega_{\pm} = \left(1\pm \gamma_5 \right) /2$ are the spin projection operators.  We find a similar 
expression when calculating the mass effects arising from $e^{-} \left(p_2 \right)$.  The explicit expressions 
for the mass correction terms $|{\cal M}^{h,b}_{ij}|^2$ can be found in Eq.~\ref{bremmass} 
in the Appendix.

Finally, we must discuss the mass effects which are relevant when $t,u,\tp$, or $\up$ becomes of ${\cal O}\left( m_{e}^{2}\right)$.  
Such terms arise only from photon exchange diagrams, and cancel 
from the numerator of $A_{LR}$.  We consequently need not worry about any complications arising from polarized 
initial beams.  Eq.~\ref{bremmass} already contains terms of the form $m_{e}^{2}c_{ij}^{2}\left( 
x\right) d_{k}^{2} \rightarrow m_{e}^{2}d_{k}^{2} / x^{2}$, which become important when either $x$ or 
$1/d_{k}$ is of ${\cal O}\left(m_{e}^{2}\right)$.  We need therefore only derive the terms of the form 
$m_{e}^{2}d_{i}d_{j} /x^{2}$.  It is possible to have ``overlapping peaks'', where both $x$ and $1/d_{i}$ become 
of ${\cal O}\left(m_{e}^{2}\right)$ at the same point in phase space.  Such events are characterized by a photon traveling 
in one direction down the beam axis while an electron goes in the opposite direction; momentum 
conservation then implies that the remaining electron has a very low energy.  The experimental cuts relevant for the
E-158 measurement require the detection of at least one electron with a reasonably high energy at a large center-of-mass frame 
angle, thereby eliminating all overlapping peaks.  The appropriate mass terms to add to the 
total unpolarized cross section, $|{\cal M}^{h,c}_{{\rm unpol}}|^2$, can be found in Eq.~\ref{unpolmass} in the Appendix; 
the same contribution enters the denominator of $A_{LR}$.

We now have all of the pieces needed for the hard bremsstrahlung cross section.  The matrix elements appearing in the 
numerator of the polarization asymmetry are 
\begin{equation}
|{\cal M}^{h}_{ij}|^2 = |{\cal M}^{h,a}_{ij}|^2 +|{\cal M}^{h,b}_{ij}|^2 \,\, ;
\label{bremeq1}
\end{equation}
the matrix elements needed for the unpolarized cross section and the asymmetry denominator become 
\begin{equation}
|{\cal M}^{h}_{{\rm unpol}}|^2 = \sum_{i,j} \left\{ |{\cal M}^{h,a}_{ij}|^2 +|{\cal M}^{h,b}_{ij}|^2 \right\} 
+  |{\cal M}^{h,c}_{{\rm unpol}}|^2 \,\, .
\label{bremeq2}
\end{equation}  

We must now present our parameterization of the $e^- e^- \rightarrow e^- e^- \gamma$ phase space.  We follow~\cite{PK} 
and write the hard bremsstrahlung cross section for a given helicity configuration as
\begin{equation}
\sigma^{h}_{ij} = \frac{1}{4s \left(2\pi\right)^{5}} \int d\sp \frac{d^{3}k}{2E_k} \frac{d^{3}q}{2E_q} 
\delta^{(4)}\left(p_1+p_2-k-q\right) \int \frac{d^{3}q_1}{2E_1}\frac{d^{3}q_2}{2E_2}\delta^{(4)}\left(q-q_1-q_2 \right)
|{\cal M}^{h}_{ij}|^2 \,\, ,
\end{equation}
where $q=q_1+q_2$ and the two sets of integrations are separately Lorentz invariant.  We choose to evaluate 
the left set in the center-of-mass (CM) frame, and the right set in the rest frame of the $q_1+q_2$ system.  Doing so, 
using $\sp = \left(q_1+q_2\right)^2 = \left(p_1+p_2-k\right)^2 = s-2E_k\sqrt{s}$, and performing an 
azimuthal integration, we arrive at 
\begin{equation}
\sigma^{h}_{ij} = \frac{1}{64s\left(2\pi\right)^{4}} \int dE_k\,dc_k\,d\Omega^{R}_{12}\,E_k\,|{\cal M}^{h}_{ij}|^2 \,\, .
\end{equation}
$\Omega^{R}_{12}$ is the rest-frame solid angle of the final-state electron $e^- \left(q_1\right)$; all rest-frame quantities 
will be denoted by a superscript $R$, while all lab (fixed target) frame quantities will have a superscript $L$.  Expressions 
without superscripts, such as the photon energy $E_k$ or the cosine of its production angle $c_k = {\rm cos}\left(\theta_k\right)$, 
denote CM frame quantities.  In the absence of experimental cuts, the limits of integration are 
\begin{eqnarray}
\Delta E \leq E_k \leq \frac{s-4 \, m_{e}^{2}}{2\sqrt{s}} \,\, , & & -1 \leq c_k \leq 1 \,\, , \nonumber \\ 
-1 \leq c^{R}_{1} = {\rm cos}\left(\theta^{R}_{1}\right) \leq 1 \,\, , & & 0 \leq \phi^{R}_1 \leq 2\pi \,\, . 
\end{eqnarray}
The relevant cuts involve restrictions on the lab frame electron energies and angles; we implement them numerically.  
It is a simple task to express the Mandelstam invariants of Eq.~\ref{mandelstam} in terms of these integration variables, 
and to derive the Lorentz transformations connecting the lab, CM, and rest frames.

The complete expressions for the unpolarized cross section and polarization asymmetry become
\begin{eqnarray}
\sigma_u &=& \frac{1}{4} \sum_{ij}\left\{ \sigma_{ij} + \sigma^{h}_{ij} \right\} \,\, , \nonumber \\ 
A_{LR} &=& \frac{\sigma_{LL}+\sigma^{h}_{LL}+\sigma_{LR}+\sigma^{h}_{LR}-\sigma_{RL}-\sigma^{h}_{RL}
- \sigma_{RR}-\sigma^{h}_{RR}}{ 4 \, \sigma_u} \,\, ,
\end{eqnarray}
where $\sigma_{ij}$ is the $2 \rightarrow 2$ cross section defined in Eq.~\ref{cross}, and the one-loop matrix 
elements $|{\cal M}_{ij}|^{2}$ are defined in Eq.~\ref{oneloop}.  We can now study the effects of both virtual 
and hard bremsstrahlung corrections on measurements of the total cross section and polarization asymmetry in 
low energy {\Moller} scattering.

\section{Numerical results}

In this section we present numerical results for the experimental setup relevant for the E-158 measurement at SLAC.  For a 
detailed description of the E-158 experiment, we refer the reader to~\cite{Carr:1997fu,E158talk}; we give below a 
description of the cuts relevant for our purposes.  We remind the reader that we have split the radiative corrections into QED 
and weak components; we will study the effects of these two types of corrections separately.

We demand that at least one electron be detected, with its lab frame energy and production angle satisfying the following 
constraints:
\begin{equation}
E^{L} \geq 10 \, {\rm GeV} \,\, , \;\;\;\;\; 4.65 \, {\rm mrad} \leq \theta^{L} \leq 8.00 \, {\rm mrad} \,\, .
\label{cuts}
\end{equation}
The exact form of the angular cut is chosen so that typical scattered electrons have a CM frame angle in the range 
$-0.5 \lsim c_1 \lsim 0$.  The structure of the E-158 detector is such that events where both scattered electrons enter 
the acceptance defined by Eq.~\ref{cuts} are counted twice; with this restriction on $\theta^{L}_{1}$, no $2 \rightarrow 2$ 
scattering process leads to such an overcounting.  However, events with hard photon emission can create such an effect; we will 
study their contribution to both the total cross section and $A_{LR}$.  We choose a beam energy $E_b = 48$ GeV, and the following 
gauge and Higgs boson masses: $M_W = 80.451$ GeV, $M_Z = 91.1875$ GeV, and $m_H = 120$ GeV.  We use the fermion masses 
and fine structure constant given in~\cite{Groom:in}.  With this choice of parameters and cuts, we find the following results 
for the Born-level and weak-corrected unpolarized cross section and polarization asymmetry:
\begin{eqnarray}
\sigma^{B}_{u} = 15.348 \, {\rm \mu b} \,\, , & & A^{B}_{LR} = 3.646 \times 10^{-7} \,\, , \nonumber \\ 
\sigma^{B+W}_{u} = 15.589 \, {\rm \mu b} \,\, , & & A^{B+W}_{LR} = 1.756 \times 10^{-7} \,\, .
\label{noQED}
\end{eqnarray}
The effect of weak radiative corrections is to decrease the polarization asymmetry by $\approx 52 \%$.  A $48\%$ 
decrease from weak corrections was found in~\cite{Denner:1998um} using the on-shell scheme; this result is for 
the specific phase space point $c_1 = 0$, different values for $M_W$, $M_Z$, and $m_H$, and an older parameterization of 
$\Delta \alpha_{{\rm had}}$.  The two results agree when these differences are taken 
into account.  We will compare our results with the $\overline{MS}$ calculation of~\cite{Czarnecki:1995fw} after 
discussing the QED corrections.  We note that the dependence of both the total cross section and $A_{LR}$ on $m_H$ is weak; 
choosing $m_H = 200$ GeV leads to an $\approx$ 3\% increase in $A_{LR}$ and a neglgible change in the cross section.

We present in Table~\ref{nowgt} the QED-corrected cross section and asymmetry, and show the percent shifts 
from the weak-corrected values $\sigma^{B+W}_{u}$ and $A^{B+W}_{LR}$ given in Eq.~\ref{noQED}.  Both virtual and hard 
bremsstrahlung corrections are included.  In our bremsstrahlung calculation we introduced an energy cutoff $\Delta E$, below 
which the emitted photon phase space integrals were performed analytically in a soft-photon approximation, and above which the 
exact bremsstrahlung matrix elements were treated numerically.  Since we integrate inclusively over the photon phase space, the 
dependence on $\Delta E$ should cancel; we include in Table~\ref{nowgt} our results for four choices of this parameter to verify 
that this indeed occurs.
\begin{table}[ht]
\begin{center}
\begin{tabular}{|c|c|c|c|c|}
\hline
$\Delta E$ $\left(\sqrt{s}\right)$ & $\sigma_{u}$ $\left({\rm \mu b}\right)$ & $\delta^{Q}\sigma_u$ & $A_{LR}$ 
$\left(10^{-7}\right)$ & $\delta^{Q}A_{LR}$  \\ \hline\hline
$10^{-6}$ & $15.455$ & $-0.86\%$ & $1.817$ & $+3.5\%$ \\ \hline
$10^{-5}$ & $15.456$ & $-0.85\%$ & $1.816$ & $+3.4\%$ \\ \hline
$10^{-4}$ & $15.457$ & $-0.85\%$ & $1.817$ & $+3.5\%$ \\ \hline
$10^{-3}$ & $15.457$ & $-0.85\%$ & $1.817$ & $+3.5\%$ \\ \hline\hline
\end{tabular}
\caption{Total unpolarized cross section, $\sigma_{u}$, and polarization asymmetry, $A_{LR}$, for different values of the 
photon energy cutoff $\Delta E$.  Also included are the shifts in these quantities from $\sigma^{B+W}_{u}$ and 
$A^{B+W}_{LR}$, defined in Eq.~\ref{noQED}, induced by QED corrections.  $\Delta E$ is given in units of $\sqrt{s}$, 
and $\sigma_{u}$ is given in millibarns.}
\label{nowgt}
\end{center}
\end{table}
These results were obtained using the Monte Carlo integration program VEGAS~\cite{Lepage:1977sw}, with the number of 
calls to the integrand $N_{call} = 5 \times 10^{6}$.  The statistical error should therefore equal $1/\sqrt{N_{call}} 
\approx 0.05\%$; we have included enough digits in the values of Table~\ref{nowgt} to test this expectation.  We see that the 
variation of $\sigma_u$ and $A_{LR}$ with $\Delta E$ is consistent with this statistical fluctuation.  The shift in the total 
cross section induced by QED corrections is small and negative.  The infrared-safe 
combination of virtual corrections and soft photon emission reduces the cross section, while the hard photon cross section 
partially compensates.  These results are similar to those of~\cite{Berends:1973jb} for Bhabha scattering, where total QED 
corrections to the cross section of $\approx -5\%$ were found for $-0.5 \leq c_1 \leq 0$, with an acollinearity cut of 
$15^{\circ}$ on the angle between the final state electron and positron and an energy cut on the detected electrons.  
Our lack of an acollinearity cut allows kinematical 
configurations in which an undetected electron travels down the beam axis, and either $u,t,\up$, or $\tp$ becomes of 
${\cal O}\left(m_{e}^{2}\right)$, adding an additional positive contribution to the QED correction.  For 
illustrative purposes we present the total cross section for three different acollinearity angles in Table~\ref{acoll}.  
The decrease in the cross section is reduced as the acollinearity angle is increased, and a larger amount of the 
photon phase space is included.

\begin{table}[ht]
\begin{center}
\begin{tabular}{|c|c|c|}
\hline
$\theta_{acoll}$ & $\sigma_{u}$ $\left({\rm \mu b}\right)$ & $\delta^{Q}\sigma_u$ \\ \hline\hline
$5^{\circ}$ & $13.865$ & $-11\%$ \\ \hline
$10^{\circ}$ & $14.407$ & $-7.6\%$ \\ \hline
$15^{\circ}$ & $14.688$ & $-5.8\%$ \\ \hline\hline
\end{tabular}
\caption{Total unpolarized cross section, $\sigma_{u}$, for different choices of the acollinearity angle $\theta_{acoll}$.  
Included are the percent shifts from the weak-corrected value in Eq.~\ref{noQED}.}
\label{acoll}
\end{center}
\end{table}

We observe from Table~\ref{nowgt} that QED corrections increase the prediction for $A_{LR}$ by $\approx 3.5\%$, leading to a final 
prediction of $A_{LR} = 1.82 \times 10^{-7}$ in the on-shell scheme.  The one-loop $\overline{MS}$ result of~\cite{Czarnecki:1995fw} 
is $A_{LR} = \left( 1.80 \pm 0.09 \pm 0.04 \right) \times 10^{-7}$, where the quoted errors arise primarily from the hadronic 
uncertainties in $\Pi^{\gamma Z}$; this value does not include either bremsstrahlung effects or experimental cuts, and is evaluated 
at the specific phase space point ${\rm cos}\left(\theta\right) = 0$ where the tree-level prediction for $A_{LR}$ is maximized.  The 
very close agreement between these results is surprising and to an extent accidental; however, we can identify the following two 
physics effects which contribute to the agreement: 1) the significant difference between the 
tree-level prediction for $A_{LR}$ in the on-shell and $\overline{MS}$ renormalization schemes caused by differing choices of 
${\rm sin}^2 \left(\theta_W \right)$ is ameliorated once virtual corrections are included, and 2) the experimental cuts 
relevant for the E-158 measurement render hard bremsstrahlung corrections a small effect.  Also, the $\approx +3.5\%$ shift induced 
by hard photon emission accidentally brings the weak-corrected value of $A_{LR}$ in Eq.~\ref{noQED} into closer 
agreement with the result of~\cite{Czarnecki:1995fw}.  QED corrections to the photon exchange component of {\Moller} scattering 
of $\approx 100\%$ are reported in~\cite{Shumeiko:1999zd}.  We believe that the large size of these results arises from the lack 
of experimental cuts in this calculation; without such restrictions, large contributions from the peaking regions discussed in the 
previous section greatly enhance the cross section.  The effects observed in a realistic experimental setup are 
significantly smaller, as our calculation demonstrates. 

We now briefly review the error on the theoretical prediction for $A_{LR}$.  The uncertainty in the one-loop result arising 
from hadronic contributions to the $\gamma Z$ vacuum polarization function has been mentioned above and discussed in the previous section; 
although large, there is hope that it will be reduced in the future~\cite{Marctalk}.  The remaining errors can be divided into 
those arising from either higher-order QED or weak corrections.  The small magnitude of the QED corrections found in this paper, 
$\approx 3.5\%$, indicate that the inclusion of ${\cal O}\left( \alpha^2 \right)$ effects are unnecessary; the experimental 
cuts protect the E-158 measurement from large QED effects.  The one-loop weak corrections are large: $\approx 40\%$ in the 
$\overline{MS}$ scheme and $\approx 50\%$ in the on-shell scheme.  However, the size of these effects arises from quark 
contributions to $A_{LR}$ not suppressed by $1-4 \, {\rm sin}^2 \left(\theta_W \right)$, a qualitatively new 
feature which first appears at ${\cal O}\left(\alpha\right)$.  No similar effect appears at the next order in perturbation 
theory, and we expect that the higher order corrections are suppressed by a factor of $\alpha / \pi \sim 0.1\%$ 
relative to the one-loop result\footnote{Certain corrections, such as those related the running of $\alpha$, will induce effects that are 
suppressed relative to the one-loop result by $\alpha /\pi \, {\rm ln}\left(M_{Z}^{2} / m_{e}^{2} \right) \approx 5\%$; however, this is 
still a small shift, and a large class of these contributions can be resummed if better precision is required.}.  We therefore conclude 
that the most significant source of theoretical uncertainty on the measurement of $A_{LR}$ arises from the hadronic contributions to 
the $\gamma Z$ vacuum polarization.

As mentioned above, the E-158 measurement double-counts events where a hard photon knocks both electrons into the 
detector acceptance; this modifies the prediction for both $\sigma_u$ and $A_{LR}$.  We find that this leads to the 
following effective values of the cross section and asymmetry, as well as the following shifts from the 
weak-corrected results of Eq.~\ref{noQED}:
\begin{eqnarray}
\sigma_{u}^{eff} = 15.613 \, {\rm \mu b} \,\, , & & \delta^{Q}_{eff}\sigma_{u} = +1.5\% \,\, , \nonumber \\
A_{LR}^{eff} = 1.829 \times 10^{-7} \,\, , & & \delta^{Q}_{eff}A_{LR} = +4.2\% \,\, .
\end{eqnarray}
We have chosen $\Delta E = 10^{-4} \sqrt{s}$ to obtain these values, and have checked that the variation with this parameter is 
consistent with the statistical error of the integration.  The overcounting of events leads to an effective increase 
of the cross section, and increases the shift of $A_{LR}$ to $+4.2\%$; subtracting from this the $\approx 3.5\%$ shift 
of Table~\ref{nowgt}, we find that the effect of overcounting is an $\approx +0.7\%$ shift in the measured value 
of $A_{LR}$.  This effect is small, however, and well within the theoretical error discussed above.

\section{Conclusions}

We have presented a calculation of the complete ${\cal O}\left(\alpha\right)$ radiative corrections to {\Moller} scattering, 
and have discussed their effect on the total cross section and polarization asymmetry measured in E-158, a low 
energy fixed target experiment being performed at SLAC.  The virtual EW corrections have been previously computed 
in~\cite{Denner:1998um}; our recalculation serves as a check of this result, with which we find complete agreement.  
Our computation of the hard bremsstrahlung contributions with realistic experimental cuts is the first such result, and 
provides a complete theoretical prediction for comparison with the E-158 measurement.  We find that hard bremsstrahlung 
effects induce a small $\approx +4\%$ shift in the measured value of $A_{LR}$; the experimental cuts reduce the 
sensitivity of both the polarization asymmetry and total cross section to QED corrections.

We have also reviewed the theoretical uncertainty in the prediction for $A_{LR}$.  The largest error arises from hadronic 
contributions to the $\gamma Z$ vacuum polarization function $\Pi^{\gamma Z}$, as discussed in~\cite{Czarnecki:1995fw}; an improved 
dispersion relation analysis of low energy $e^+ e^- \rightarrow \, {\rm hadrons}$ data could reduce this value~\cite{Marctalk}.  
Our results indicate that higher order QED corrections are likely to be small in comparison to the error on 
$\Pi^{\gamma Z}$, and can be safely neglected in the E-158 analysis.  The one-loop weak corrections are $\approx 50\%$; although 
this effect seems dangerously large, it arises from the appearance of effects unsuppressed by the small electron  
vector coupling, a qualitatively new contribution which first enters at one loop.  As no similar effect appears at higher 
orders, we conclude that the perturbative corrections to $A_{LR}$ are under control.

The results obtained here are applicable to both low energy experiments such as E-158 and future high energy colliders, 
and provide detailed Standard Model predictions to which measured values can be compared.  We hope that they 
assist in maximizing the discovery potential of both current and future experimental programs. 

\noindent{\Large\bf Acknowledgments}

\noindent
It is a pleasure to thank K. Melnikov for suggesting this project, and for his constant encouragement and assistance.  I would also like to 
thank C. Anastasiou, P. Bosted, L. Dixon, J. Hewett, Y. Kolomensky, T. Rizzo and M. Woods for helpful comments and suggestions.  This work 
was supported in part by the National Science Foundation Graduate Research Program.

\noindent{\Large\bf Appendix}

\noindent
We collect here the expressions for the various hard bremsstrahlung matrix elements 
$|{\cal M}^{h,x}_{ij}|^2$ discussed in Eqs.~\ref{bremeq1} and~\ref{bremeq2}.  We first consider $|{\cal M}^{h,a}_{ij}|^2$, 
which are computed with $m_e = 0$
everywhere; with the Mandelstam invariants presented in Eq.~\ref{mandelstam}, $c_{ij}\left(x\right)$ 
defined in Eq.~\ref{effcoup}, and $d_{i}$ defined in Eq.~\ref{bremdenoms}, we can express them as
\begin{eqnarray}
|{\cal M}^{h,a}_{LL}|^2 &=& 8 e^2 \, \bigg\{\,  \left(s^2+s^{'2}\right)\left\{t \, c_{LL}^2\left( t \right) d_2 d_4 
+ u \, c_{LL}^2 \left( u \right) d_2 d_3 + \tp c_{LL}^2 \left( tp \right) d_1 d_3
+ \up c_{LL}^2 \left( up \right) d_1 d_4 \right\}  \nonumber \\ & & 
-2 \sp \left\{ c_{LL}\left( \tp \right) c_{LL} \left( \up \right) \left(s+\tp+\up \right) d_1 
+ c_{LL}\left( t \right) c_{LL} \left( u \right) \left(s+t+u \right) d_2 \right\} \nonumber \\ & &
-2 s \left\{ c_{LL}\left( \tp \right) c_{LL} \left( u \right) \left(\sp+\tp+u \right) d_3 
+ c_{LL}\left( t \right) c_{LL} \left( \up \right) \left(\sp+t+\up \right) d_4 \right\} \nonumber \\ & &
-2 \left\{ c_{LL}\left( \tp \right)+ c_{LL}\left( \up \right) \right\} \left\{ c_{LL}\left( t 
\right)+ c_{LL}\left( u \right) \right\}\sp \left(\tp+\up\right) \left(t+u\right) d_1 d_2 \nonumber \\ & &
-2 \left\{ c_{LL}\left( \tp \right)+ c_{LL}\left( u \right) \right\} \left\{ c_{LL}\left( t 
\right)+ c_{LL}\left( \up \right) \right\}s \left(\tp+u\right) \left(t+\up\right) d_3 d_4 \nonumber \\ & & 
-\left\{c_{LL}\left( u \right) c_{LL}\left( \tp \right) + c_{LL}\left( \tp \right) c_{LL}\left( \up \right)
+c_{LL}\left( u \right) c_{LL}\left( \up \right) \right\} \left\{ t\tp \left(s+\sp \right) \right. \nonumber \\ & & 
\left.+s\sp \left(t-\tp \right) -s\sp \left( u+\up \right) -u\up \left( s+\sp \right) \right\} d_1 d_3 
\nonumber \\ & & 
-\left\{c_{LL}\left( t \right) c_{LL}\left( \up \right) + c_{LL}\left( \up \right) c_{LL}\left( \tp \right)
+c_{LL}\left( t \right) c_{LL}\left( \tp \right) \right\} \left\{ u\up \left(s+\sp \right) \right. \nonumber \\ & & 
\left.+s\sp \left(u-\up \right) -s\sp \left( t+\tp \right) -t\tp \left( s+\sp \right) \right\} d_1 d_4 
\nonumber \\ & & 
-\left\{c_{LL}\left( u \right) c_{LL}\left( \tp \right) + c_{LL}\left( t \right) c_{LL}\left( u \right)
+c_{LL}\left( t \right) c_{LL}\left( \tp \right) \right\} \left\{ u\up \left(s+\sp \right) \right. \nonumber \\ & & 
\left.+s\sp \left( \up-u \right) -s\sp \left( t+\tp \right) -t\tp \left( s+\sp \right) \right\} d_2 d_3 
 \nonumber \\ & & 
-\left\{c_{LL}\left( t \right) c_{LL}\left( \up \right) + c_{LL}\left( u \right) c_{LL}\left( t \right)
+c_{LL}\left( u \right) c_{LL}\left( \up \right) \right\} \left\{ t\tp \left(s+\sp \right) \right. \nonumber \\ & & 
\left.+s\sp \left( \tp-t \right) -s\sp \left( u+\up \right) -u\up \left( s+\sp \right) \right\} d_2 d_4 \, \bigg\}
\,\, , \nonumber \\ 
|{\cal M}^{h,a}_{RR}|^2 &=& |{\cal M}^{h,a}_{LL} \left( c_{LL} \left( x \right) \rightarrow c_{RR} \left( x \right) 
\right) |^2 \,\, , \nonumber \\ 
|{\cal M}^{h,a}_{LR}|^2 &=& |{\cal M}^{h,a}_{RL}|^2 = 8 e^2 \, \bigg\{\, u\, c_{LR}^{2} \left( u \right) 
\left(t^2+t^{'2}\right)d_2 d_3 + t\, c_{LR}^{2} \left( t \right) \left(u^2+u^{'2}\right)d_2 d_4 \nonumber \\ & & 
+\tp c_{LR}^{2}\left(\tp\right)\left\{s^2+s^{'2}+t^2-t^{'2}-2\tp u -2\tp\up+2st+2s\sp+2\sp t \right\}d_1 d_3 
\nonumber \\ & & 
+\up c_{LR}^{2}\left(\up\right)\left\{s^2+s^{'2}+u^2-u^{'2}-2\up t -2\up\tp+2su+2s\sp+2\sp u \right\}d_1 d_4
\nonumber \\ & & 
+\left\{c_{LR}\left(t\right)c_{LR}\left(\tp\right)\left\{\left(u+\up\right)\left(s\sp-t\tp+u\up\right)+2su\up\right\}
  \right. \nonumber \\ & & \left. +c_{LR}\left(u\right)c_{LR}\left(\up\right)\left\{\left(t+\tp\right)
\left(s\sp-u\up+t\tp\right)+2st\tp\right\}\right\}d_1 d_2
\nonumber \\ & & 
+\left\{c_{LR}\left(t\right)c_{LR}\left(\tp\right)\left\{\left(u+\up\right)\left(s\sp-t\tp+u\up\right)+2\sp u\up\right\}
  \right. \nonumber \\ & & \left. +c_{LR}\left(u\right)c_{LR}\left(\up\right)\left\{\left(t+\tp\right)
\left(s\sp-u\up+t\tp\right)+2\sp t\tp\right\}\right\}d_3 d_4
\nonumber \\ & & 
+2 c_{LR}\left(u\right)c_{LR}\left(\up\right)\tp\left(s+u\right)\left(\sp+\up\right)d_1 d_3 \nonumber \\ & & 
+2 c_{LR}\left(t\right)c_{LR}\left(\tp\right)\up\left(s+t\right)\left(\sp+\tp\right)d_1 d_4 \nonumber \\ & & 
+2 c_{LR}\left(t\right)c_{LR}\left(\tp\right)u\left(s+\tp\right)\left(\sp+t\right)d_2 d_3 \nonumber \\ & & 
+2 c_{LR}\left(u\right)c_{LR}\left(\up\right)t\left(s+\up\right)\left(\sp+u\right)d_2 d_4  \, \bigg\} \,\, .
\label{naive}
\end{eqnarray}
We now present $|{\cal M}^{h,b}_{ij}|^2$, the mass terms which become important when $1/d_{i} \rightarrow 
{\cal O}\left(m_{e}^{2}\right)$; we find
\begin{eqnarray}
|{\cal M}^{h,b}_{LL}|^2 &=& 8 e^2 \frac{m_{e}^{2}}{s} \, \bigg\{ \, -\left\{c_{LL}\left(\tp\right)+
c_{LL}\left(\up\right)\right\}^{2} \sp\left\{s^2+\left(\tp+\up\right)^{2}\right\}d_{1}^{2} 
+c_{LR}^{2}\left(\tp\right) \up \left\{\tp \right. \nonumber \\ & &  \left. \times \left(s-u-t\right) 
+su+2s\sp+\sp\up\right\} d_{1}^{2} 
+ c_{LR}^{2}\left(\up\right) \tp \left\{\up \left(s-u-t\right)  \right. \nonumber \\ & &  \left. 
+st+2s\sp+\sp\tp\right\} d_{1}^{2} 
-\left\{c_{LL}\left( t\right)+
c_{LL}\left( u\right)\right\}^{2} \sp\left\{s^2+\left( t+u\right)^{2}\right\}d_{2}^{2} 
\nonumber \\ & &  +c_{LR}^{2}\left( t\right) u \left\{t \left(s-\up-\tp\right) 
+s\up+2s\sp+\sp u\right\} d_{2}^{2} 
+ c_{LR}^{2}\left( u\right) t \left\{u \left(s-\up-\tp\right)  \right. \nonumber \\ & &  \left. 
+s\tp+2s\sp+\sp t\right\} d_{2}^{2} +2 \left\{c_{LL}\left( u \right)+c_{LL}\left(\tp \right)\right\}^{2} 
s^2 \left(u+\tp\right)d_{3}^{2} \nonumber \\ & & 
+2 \left\{c_{LL}\left( t \right)+c_{LL}\left(\up \right)\right\}^{2} 
s^2 \left(t+\up\right)d_{4}^{2} \, \bigg\} \,\, , \nonumber \\ 
|{\cal M}^{h,b}_{RR}|^2 &=& |{\cal M}^{h,b}_{LL}\left(c_{LL}\left(x\right) \rightarrow c_{RR}
\left(x\right)\right)|^2 \,\, , \nonumber \\
|{\cal M}^{h,b}_{LR}|^2 &=& 8 e^2 \frac{m_{e}^{2}}{s} \, \bigg\{ \, \left\{c_{RR}\left(\tp\right)+
c_{RR}\left(\up\right)\right\}^{2}\sp\left(s+\tp+\up\right)^{2}d_{1}^{2} 
-c_{LR}^{2}\left(\tp\right) \up \nonumber \\ & & \times \left\{\tp\left(\sp+\tp+\up\right)
+su+\sp\up\right\}d_{1}^{2}-c_{LR}^{2}\left(\up\right)\tp\left\{\up\left(\sp+\tp+\up\right)
\right. \nonumber \\ & & \left. +st+\sp\tp\right\}d_{1}^{2} 
+ \left\{c_{LL}\left( t\right)+
c_{LL}\left( u\right)\right\}^{2}\sp\left(s+t+u\right)^{2}d_{2}^{2} 
-c_{LR}^{2}\left( t\right) u \nonumber \\ & & \times \left\{ t\left(\sp+t+u\right)
+s\up+\sp u\right\}d_{2}^{2}-c_{LR}^{2}\left( u\right) t\left\{ u\left(\sp+t+u\right)
\right. \nonumber \\ & & \left. +s\tp+\sp t\right\}d_{2}^{2} +2\,c_{LR}^{2}\left( u\right)\tp\left( s+u\right)d_{3}^{2}
+2\,c_{LR}^{2}\left(\tp\right) u\left( s+\tp\right)d_{3}^{2} \nonumber \\ & & 
 +2\,c_{LR}^{2}\left( t\right)\up\left( s+t\right)d_{4}^{2} 
+2\,c_{LR}^{2}\left(\up\right) t\left( s+\up\right)d_{4}^{2} \,\bigg\} \,\, , \nonumber \\ 
|{\cal M}^{h,b}_{RL}|^2 &=& |{\cal M}^{h,b}_{LR}\left(c_{LL}\left(x\right) \rightarrow c_{RR}
\left(x\right)\right)|^2 \,\, .
\label{bremmass}
\end{eqnarray}
Finally, the modification of the total unpolarized cross section that results when $t,u,\tp,\up \rightarrow 
{\cal O}\left(m_{e}^{2}\right)$ is 
\begin{eqnarray}
|{\cal M}^{h,c}_{{\rm unpol}}|^2 &=& 64 \,e^{6}\,m_{e}^{2} \left\{ \frac{\left(s+u\right)\left(\sp+\up\right)d_2 d_4}{t^2}
+ \frac{\left(s+t\right)\left(\sp+\tp\right)d_2 d_3}{u^2} \right. \nonumber \\ & & \left. + \frac{\left(s+\up\right)
\left(\sp+u\right)d_1 d_3}{t^{'2}} + \frac{\left(s+\tp\right)\left(\sp+t\right)d_1 d_4}{u^{'2}} \right \} \,\, .
\label{unpolmass}
\end{eqnarray}

\end{document}